\documentclass{Interspeech2024}

\interspeechcameraready

\usepackage{multirow}
\usepackage{diagbox}
\usepackage[capitalize]{cleveref}
\usepackage{makecell}

\usepackage[natbib, bibencoding=utf8, citestyle=numeric, bibstyle=ieee, maxbibnames=999, maxcitenames=2, mincitenames=1, sortcites]{biblatex}
\bibliography{mybib}

\usepackage{todonotes}

\usepackage[acronym, shortcuts, nohypertypes={acronym}]{glossaries}
\newacronym{CLAP}{CLAP}{contrastive language-audio pretraining}
\newacronym{CASA}{CASA}{computational audiotiry scene analysis}
\newacronym{CP}{CP}{computational paralinguistics}
\newacronym{LLM}{LLM}{large language model}
\newacronym{ML}{ML}{machine learning}
\newacronym{SER}{SER}{speech emotion recognition}

\newcommand{\eg}{e.\,g.,\,}

\newcommand{\eclap}{ParaCLAP }

\newcommand\blfootnote[1]{%
  \begingroup
  \renewcommand\thefootnote{}\footnote{#1}%
  \addtocounter{footnote}{-1}%
  \endgroup
}

\title{ParaCLAP -- Towards a general language-audio model \\for computational paralinguistic tasks}
\name[affiliation={*\,1,2}]{Xin}{Jing}
\name[affiliation={*\,1,2}]{Andreas}{Triantafyllopoulos}
\name[affiliation={1,2,3}]{Björn}{Schuller}

\address{
  $^1$Chair of Embedded Intelligence for Health Care \& Wellbeing, University of Augsburg, Germany\\
  $^2$CHI -- Chair of Health Informatics, MRI, Technical University of Munich, Germany \\
  $^3$GLAM -- Group on Language, Audio, \& Music, Imperial College London, UK}
\email{xin.jing@tum.de}

\keywords{computational paralinguistic, speech emotion recognition, contrastive learning, zero-shot learning}

\begin{document}

\maketitle
 
\begin{abstract}
\ifinterspeechfinal
\blfootnote{* These authors contributed equally}
\fi
Contrastive language-audio pretraining (CLAP) has recently emerged as a method for making audio analysis more generalisable.
Specifically, CLAP-style models are able to `answer' a diverse set of language queries, extending the capabilities of audio models beyond a closed set of labels.
However, CLAP relies on a large set of (audio, query) pairs for pretraining.
While such sets are available for general audio tasks, like captioning or sound event detection, there are no datasets with matched audio and text queries for computational paralinguistic (CP) tasks.
As a result, the community relies on generic CLAP models trained for general audio with limited success.
In the present 
study, 
 we explore training considerations for ParaCLAP, a CLAP-style model suited to CP, including a novel process for creating audio-language queries.
We demonstrate its effectiveness on a set of computational paralinguistic tasks, where it is shown to surpass the performance of open-source state-of-the-art models.

\end{abstract}

\section{Introduction}
\Ac{CP} is a subfield of affective computing which corresponds to the analysis of the paralinguistic information in speech signals for the prediction of speaker states and traits~\citep{Schuller14-CPE}.
As such, it encompasses a large gamut of phenomena that can be predicted: from the now `classic' emotions, to (among many others) personality, likability, sincerity, deception, and even health-related tasks.
The typical \ac{ML} workflow requires the collection of representative data for each of those tasks and the subsequent training of models -- oftentimes relying on transfer learning or unsupervised pretraining to improve performance while reducing the dependency on data quantity~\citep{wagner2023dawn}.
Nonetheless, the challenge remains that at least \emph{some} amount of data has to be acquired for each task, which presents a bottleneck for practitioners who want to benefit from advances in \ac{CP} but are interested in phenomena where no existing data or public models exist.

A similar challenge is faced by the \ac{CASA} community, which also aims to analyse a wide spectrum of scenes and events~\citep{Drossos20-CAA, Xie21-ZSA, Mei22-AAC}.
However, that field has seen the recent emerge of \ac{CLAP} as a general-purpose method which can overcome this lack of data and result in a generic model that can be employed for a wide variety of tasks.
The initial inspiration for CLAP came from computer vision~\citep{Radford21-LTV}.
\ac{CLAP} is trained to compute the similarity of linguistic and audio embeddings (each generated by the respective linguistic/auditory encoders)~\citep{Elizalde23-CLA}.
During inference, \ac{CLAP} relies on a set of linguistic queries (provided by the user) which are scored with respect to their compatibility with the input audio.
Importantly, this allows the application of the model on \emph{completely new} tasks never seen during training.
On the downside, its success is heavily based on the large amount of pretraining data formed of (audio, query) pairs.
In the \ac{CASA} domain, the community has amassed a large set of data that can be used for this purpose, \eg by using audio captioning datasets~\citep{Drossos20-CAA, Mei22-AAC} or by co-opting the tags and descriptions associated with an audio sample in large data platforms like Freesound~\citep{Fonseca17-FDA}.
As no such public data exist for \ac{CP} tasks, prior research has primarily used \ac{CP} tasks in downstream evaluations of zero-shot performance.
For example, \citet{Elizalde23-CLA} evaluated their \ac{CLAP} model on two \ac{SER} datasets and obtained results slightly above chance-level performance.
This shows how general \ac{CLAP} models pretrained on \ac{CASA} tasks may struggle in the \ac{CP} domain.

On a related, more promising note, a similar setup has been recently used to improve the expressivity of affective speech synthesis.
For example, PromptTTS-2\citep{leng2023prompttts} constructed a set of linguistic ``prompts'' used to control their synthesis system.
This set is created using templates derived from acoustic parameters and showed improved effectiveness in synthesising the target attributes.
While such approaches illustrate the potential of generating linguistic prompts suitable to describe paralinguistic attributes, to the best of our knowledge they have not been used to improve zero-shot recognition for \ac{CP} tasks.
Moreover, previous work on zero-shot learning has shown that this is possible for \ac{SER}~\citep{Xu21-RAA}, and in particular using semantic embeddings derived directly from the labels~\citep{Xu21-EZS}.
Similarly, these findings have not been extended to a more broader range of \ac{CP} tasks.

Our work aims to bridge this gap by presenting a novel templating method based on interpretable, expert features known to capture paralinguistic attributes.
Starting from those features and the set of labels available for a large \ac{SER} dataset (MSP-Podcast), we construct a diverse set of (audio, query) samples.
These pairs are given as input to an audio (wav2vec2.0) and text (BERT) encoder, which procure the corresponding embeddings.
During training, these embeddings are \emph{aligned} using a contrastive loss.
During inference, the similarity of the input audio and the target linguistic queries is computed, and the query with the highest similarity is assigned as the predicted class.
We demonstrate the effectiveness of our method on a set of \ac{SER} datasets to first test their out-of-domain performance, as well as two broader \ac{CP} tasks to gauge their generalisability.

The remainder of this paper is organised as follows:
Our proposed templating process and model are introduced in \cref{sec:experiment}.
\cref{sec:result} presents our results and analysis.
Finally, conclusion and future work are discussed in \cref{sec:conclusion}.


\section{Methodology}
\label{sec:methodology}
\subsection{\eclap}

\begin{figure}[t]
    \centering
    \includegraphics[width=0.48\textwidth]{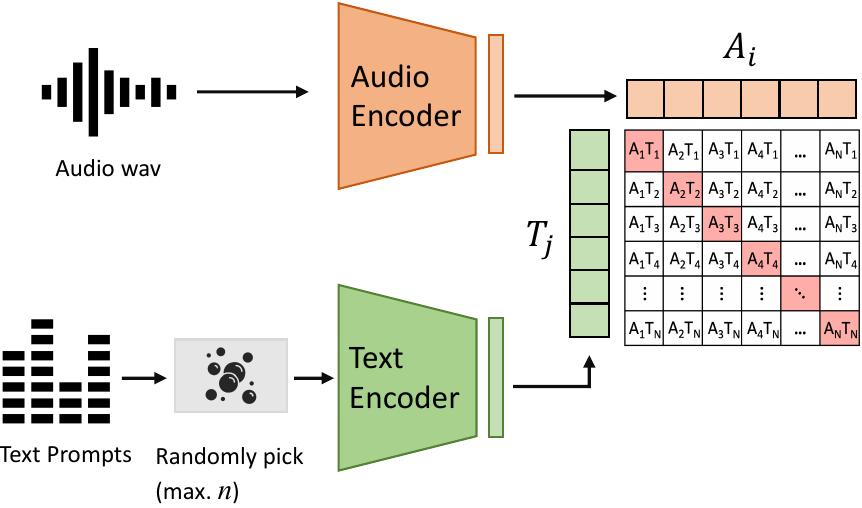}
    \caption{Diagram of our proposed \eclap model. \eclap jointly trains two parallel encoders to establish a link between the input audio-text pairs in a shared multimodal space. In the training phase, a maximum of $n$ text prompts are randomly selected from the pool of generated text queries, combining both label and pseudo-caption information, while during inference, the query comprises only the actual labels.}
    \vspace{-0.6cm}
    \label{fig:e-clap}
\end{figure}

Our overall workflow follows the standard procedure from \ac{CLAP} models~\citep{Elizalde23-CLA}.
As shown in \cref{fig:e-clap}, \eclap establishes a link between audio and text representations within a shared multimodal space by applying two encoders and aligning their output embeddings using
contrastive learning. 

In detail, let $\{X^a_i, X^t_i\}$ be the input audio-text pairs, in which $i \in [0, N]$ and $N$ is the batch size.
The audio-text pairs will be processed by an audio and text encoder, as shown in \cref{fig:e-clap}, to extract an audio and a text embedding. 
Additionally, every encoder incorporates a projection layer which serves the purpose of projecting both the audio and text embeddings into a shared feature space of dimensionality $d$. Every projection layer is a module consisting of two linear transformations followed by a GELU activation and layer normalisation. It initially maps the input to larger dimensions and then downsamples to its original dimensions, preserving a consistent structure.
In our study, the output size of the projection layer is $768$. 
The final audio and text embeddings are computed as $A_i=proj_A(f(X^a_i)), A_i \in \mathbb{R}^{N\times d}$, and $T_i=proj_T(g(X^t_i)), T_i \in \mathbb{R}^{N\times d}$, where the $f(\cdot)$ and $g(\cdot)$ are the audio encoder and text encoder, respectively.
In this shared feature space, the text embedding $T_i$ and audio embedding $A_i$ can be compared by measuring their similarity.
For this purpose, we use the scaled cosine similarity:
\begin{equation}
    S = \tau * (A_i \cdot T_i^\mathsf{tr}), 
\end{equation}
where $\tau$ is a temperature parameter to scale the range of log-
its.
The similarity matrix $Sim$ contains $N$ positive pairs in the diagonal and $N^2 - N$ negative pairs in the off-diagonal.

To compute the final loss for training, the following symmetric cross-entropy loss is applied:
\begin{equation}
    Loss = 0.5 * (h_{t}(S) + h_{a}(S)),
\end{equation}
 where the $h_{k} = \frac{1}{N}\sum^{N}_{n=0}log(diag(softmax(Sim)))$ along the audio and text axis.
The CLAP model is thus optimised by maximising the normalised similarity of positive pairs and minimising the similarity of negative pairs, which means 
the 
asymptotic limit of diagonal elements approaching 1 ($\lim_{{i \to \infty}} \text{{diag}}(S)_{ii} = 1$), and non-diagonal elements converging towards zero ($\lim_{{i \neq j \to \infty}} \text{{off}}(S)_{ij} = 0$).

In this study, we adopt the wav2vec 2.0 large model \citep{baevski2020wav2vec} as the audio encoder, initialised with a pre-trained state which has been finetuned for dimensional \ac{SER}\footnote{https://huggingface.co/audeering/wav2vec2-large-robust-12-ft-emotion-msp-dim}\citep{wagner2023dawn}. 
The pre-trained model was pruned from 24 to 12 transformer layers before fine-tuning on the MSP-Podcast (v1.7) dataset\citep{lotfian2017building}. The pre-trained wav2vec 2.0 large model consists of 160 million parameters.
The audio embedding is obtained by extracting the pooled states from the last transformer layer which has a size of 1024. For the text encoder, we apply the BERT\citep{devlin2018bert} base model (uncased)\footnote{https://huggingface.co/bert-base-uncased} implemented by HuggingFace\citep{wolf2019huggingface}. The model has 110 million parameters and the [CLS] token extracted from the final layer is utilised as the text embedding. 

\subsection{Query generation}
\label{ssec:query}
We generate text queries matching each sample by relying on two main sources of information:
a) The labels already included in the dataset.
b) Expert acoustic and prosodic features extracted for each sample.
Each source of query is extended to a complete sentence as described below.
A comprehensive table is provided as supplementary material\footnote{The process can be more easily understood by inspecting the code submitted as supplementary material.}.

\noindent
\textbf{Dataset labels:}
Our training dataset (\cref{ssec:training}) is labelled for emotion, both categorical and dimensional (arousal, valence, dominance), as well as gender.
Categorical emotion is expanded by constructing sentences like ``this is a [EMOTION] instance" or c) ``speaker is [EMOTION]", with [EMOTION] coming in the form of adjective (\eg ``angry", ``happy", etc.).
The gender label (``male" is expanded in two versions:
a) a [GENDER] is speaking; b) the speaker is [GENDER].
Finally, the dimensional emotional attributes are first binned according to their distribution (bottom 30\,\%, middle 40\,\%, top 30\,\%) and templates are constructed accordingly (\eg arousal is low/mid/high).

\noindent
\textbf{Pseudo-captions using expert features:}
We extract the extended Geneva Minimalistic Acoustic Parameter Set (eGeMAPS)~\citep{Eyben15-TGM} using openSMILE~\citep{Eyben10-OTM}. 
It includes 88 parameters and is known for its effectiveness in several \ac{CP} tasks. 
In this study, we focus on the following features which are easier to interpret:
a) mean ($\mu$) and standard deviation ($\sigma$) of \emph{pitch};
b) sound \emph{intensity} (over whole utterance);
c) \emph{jitter} (mean over utterance);
d) \emph{shimmer} (mean over utterance).
In addition, we compute the total \emph{duration} of each utterance.
This gives us 5 numerical features, which we proceed to bin according to their distribution as for the dimensional variables.
Following that, we generate queries using those bins (\eg ``pitch is low/normal/high").

\noindent
\textbf{Combinations:}
Finally, during training we combine several queries at once using an ``and" conjunction (\eg ``speaker is happy and pitch is high").
This further expands the diversity of data that the model sees during training, and is aimed to improve training performance and prevent overfitting.

\begin{table*}[!h]
\vspace*{-8mm}
\centering
\caption{The Unweighted Average Recall (UAR) results on 7 datasets, comprising both English and German language-based datasets.  `emo' indicates the use of emotion queries during training. The notation $rand=n$ means a maximum of $n$ queries are randomly selected and concatenated. If `emo' is mentioned, it implies emotion queries' involvement in the data. \textbf{Bold} marks the best performance while \underline{underline} represents the second best.}
\begin{tabular}{l c| c |c c c c c c}
\hline
\multirow{2}{*}{\rule{0pt}{4.7ex}Dataset} & \multirow{2}{*}{\rule{0pt}{4.7ex}Class}&\multirow{2}{*}{\rule{0pt}{4.7ex}Lang.}& \multicolumn{5}{c}{\rule[-1.2ex]{0pt}{3.6ex}Unweighted Average Recall (UAR)} \\ \cline{4-9}
&&&\thead{CLAP\\(CNN14 + Bert)} & \thead{Pengi \\(HTSAT + GPT2)}&\thead{\eclap\\(no emo, $rand=5$)}& \thead{\eclap\\(only emo)} &\thead{\eclap\\($rand=1$)} & \thead{\eclap\\($rand=5$)} \\
\hline
\multicolumn{1}{l}{\rule{0pt}{2.8ex}IEMOCAP} &4cl& en & .353 &.345 & .309& \textbf{.567} & .307 & \underline{.560}\\
\midrule

\multicolumn{1}{l}{RAVDESS} &8cl& en &.199 & .148 &.170&  \textbf{.302} & .116 & \underline{.234} \\
\midrule

\multicolumn{1}{l}{CREMA-D} &6cl& en &.230 & .245 &.201&  \textbf{.332} & .202 & \underline{.291} \\
\midrule

\multicolumn{1}{l}{TESS} &7cl& en &.232 & .177 &.212&\textbf{.484}& .219 & \underline{.389} \\
\midrule

\multicolumn{1}{l}{FAU Aibo} &2cl& de &.500 & .470 &\underline{.538}&  .535 & .468 & \textbf{.604} \\
\midrule

\multicolumn{1}{l}{FAU Aibo} &5cl& de &.211 & .185 &\underline{.225}& .216 & .216 & \textbf{.232} \\
\midrule

\multicolumn{1}{l}{ALC} &2cl& de &\underline{.511} & .473 &.490&  \textbf{.512} & .501 &  .503 \\
\midrule

\multicolumn{1}{l}{SLD} &2cl& de &.472 & .485 &.472&  \textbf{.554} & .443 & \underline{.507} \\

\bottomrule
\end{tabular}
\label{tab:emods}
\end{table*}

\section{Experiments}
\label{sec:experiment}

\subsection{Training Dataset}
\label{ssec:training} 
In this work, MSP-podcast Release 1.9\,\citep{lotfian2017building} is utilised to train our \eclap models.
The MSP-podcast data comprises natural English speeches extracted from podcast recordings.
The dataset encompasses 55\,283 utterances spoken by over 1\,200 speakers, totalling more than 110 hours of speech. 
The corpus is annotated using crowd-sourcing with 9 emotion types (\textbf{neutral, fear, sadness, disgust, happiness, other, anger, contempt, surprise}), but nearly 17.5\,\% of the data does not have majority agreed labels (labelled as 'no\_agreement').
Moreover, the data is heavily imbalanced towards the neutral class ($35\%$), followed by happiness ($21\%$) and anger ($6\%$).
During the training phase, the standard splits for training and testing are employed.
However, data labelled as `no\_agreement' is excluded, resulting in a final amount of 45\,619 utterances for training.
To construct the audio-text pairs for training, we dynamically pick texts from the text queries outlined in \cref{ssec:query}, and additional details for text-audio pairs will be presented in \cref{ssec:exp_set}.

\subsection{Test Datasets}
\noindent
\textbf{IEMOCAP:} The Interactive Emotional Dyadic Motion Capture~\citep{busso2008iemocap} is a prominent standard multi-modal database for emotion studies. We follow the most frequently used category selection (angry, happy+excited, sad, and neutral) to build the test dataset. Thus, the dataset contains 5\,531 utterances (\textbf{1\,103 angry, 1\,636 happy, 1\,708 neutral}, and \textbf{1\,084 sad}).

\noindent
\textbf{RAVDESS:} The Ryerson Audio-Visual Database of Emotional Speech and Song~\citep{livingstone2018ryerson} is a multimodal emotion database containing speech and songs. The speech data consists of 1\,440 utterances with 8 expressions (\textbf{neutral, calm, happy, sad, angry, fearful, surprise}, and \textbf{disgust}).

\noindent
\textbf{CREMA-D:} Crowd Sourced Emotional Multimodal Actors Dataset~\citep{cao2014crema} is a crowd-sourced audio-visual data set tailored for emotion studies. It comprises 7\,442 clips from 91 actors. The dataset consists of facial and vocal emotional expressions in sentences, conveying a diverse range of basic emotional states (\textbf{happy, sad, anger, fear, disgust}, and \textbf{neutral}).

\noindent
\textbf{TESS:} Toronto Emotional Speech Set~\citep{pich2020tess} is an audio emotion data set recorded by two female speakers, aged 26 and 64\,years. It contains 2\,800 clips featuring seven emotions:\textbf{ anger, disgust, fear, happiness, pleasant surprise, sadness}, and \textbf{neutral}. 

\noindent
\textbf{FAU-Aibo:} The FAU-Aibo Emotion Corpus~\citep{batliner2008releasing} is a German speech emotion database containing children speech in the ages 6 to 10\,years. 
It is used in the INTERSPEECH 2009 Emotion Challenge~\citep{schuller2009interspeech}, and includes a training set of 9\,959 speech chunks and a test set of 8\,257 chunks. 
The original 11 labels are mapped to
a) a five-category classification problem, with labels being merged into \textbf{angry, emphatic, neutral, positive}, and \textbf{rest}, and
b) a two-category classification problem, with \textbf{`non-negative'} and \textbf{`negative'} as the emotion labels.

\noindent
\textbf{ALC:} The Alcohol Language Corpus~\citep{schiel2012alcohol} contains German speech collected under a systematic intoxication test. The
INTERSPEECH 2011 Speaker State Challenge~\citep{schuller2011interspeech} selected part of the ALC to obtain a gender and age balanced dataset. 1\,620 data with the label \textbf{`not intoxicated with alcohol'} and 1\,380 labelled as \textbf{`intoxicated with alcohol'} are contained in the test set.

\noindent
\textbf{SLD:} The Speaker Likability Database is a subset of the German telephone speech dataset aGender~\citep{burkhardt2010database}. It was used in the 
INTERSPEECH 2012 Speaker Trait Challenge~\citep{schuller2012interspeech} and contains 800 audio chunks each for the labels \textbf{`likable'} and \textbf{`non-likable'}.

\subsection{Experimental setup}
\label{ssec:exp_set}
We use the raw waveform after resampling to 16\,kHz for training and testing.
During training, all audio sequences are randomly clipped or padded to achieve a consistent 5-second duration and a maximum of $n$ text prompts are randomly 
selected from the pool of the text queries and concatenated into a single unit.
We additionally control whether the emotion label is part of the training query, resulting in a total of four alternatives:
\begin{itemize}
    \item \eclap(no emo, $rand=5$): emotion query is not part of training, a maximum of 5 random queries concatenated
    \item \eclap(only emo): only emotion query is used as caption
    \item \eclap($rand=1$): emotion query is part of training, a maximum of 1 random queries
    \item \eclap($rand=5$): emotion query is part of training, a maximum of 5 random queries
\end{itemize}

All models are trained with an Adam optimiser and a batch size of 64. The number of training epochs is set to 50. 
Both the audio and text branches remain unfrozen, utilising a learning rate of $1e-5$. 
Meanwhile, the projection layers and other parameters employ a learning rate set of $1e-3$. 
The models are implemented with \textsc{PyTorch-v1.13.1} on a single \textsc{Nvidia A40, 48\,GB} GPU.

The \eclap model from the epoch that yielded the best performance on the MSP-Podcast test set is selected as the best-performing model and used for the downstream evaluations.
During inference, we use the annotated labels provided by the datasets as the text query.
As baselines, we use the original CLAP model\footnote{https://huggingface.co/microsoft/msclap}, which employs a CNN14 audio encoder and a BERT text encoder, and has been trained on larger data featuring a wider gamut of sound tasks, and Pengi~\citep{Deshmukh24-PAA}, a large audio-language model trained for a variety of tasks.
Pengi generates free-form text which can be matched to each query by computing the similarity of query and output embeddings, as in the original work~\citep{Deshmukh24-PAA}.
\section{Results \& Discussion}
\label{sec:result}

\begin{figure}
    \centering
    \hspace*{-6mm}
    \includegraphics[width=0.48\textwidth]{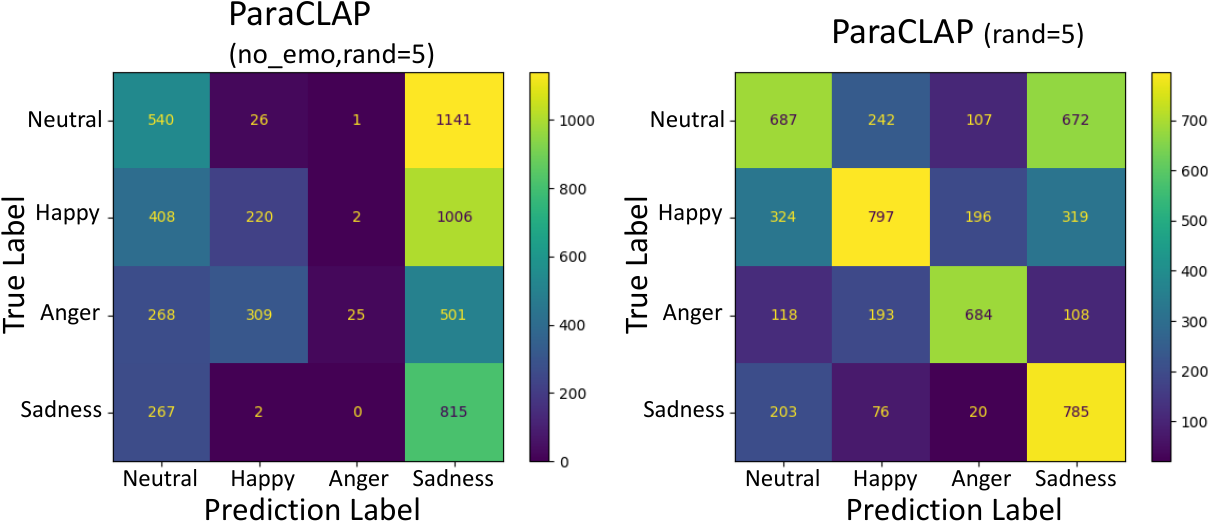}
    \caption{The confusion matrices illustrate the results on the IEMOCAP dataset. Models share identical training settings, differing in the inclusion of emotion queries within the audio-text pairs (left: without, right: with)}
    \label{fig:cm-iemo}
    \vspace*{2mm}
    \hspace*{-6mm}
    \includegraphics[width=0.48\textwidth]{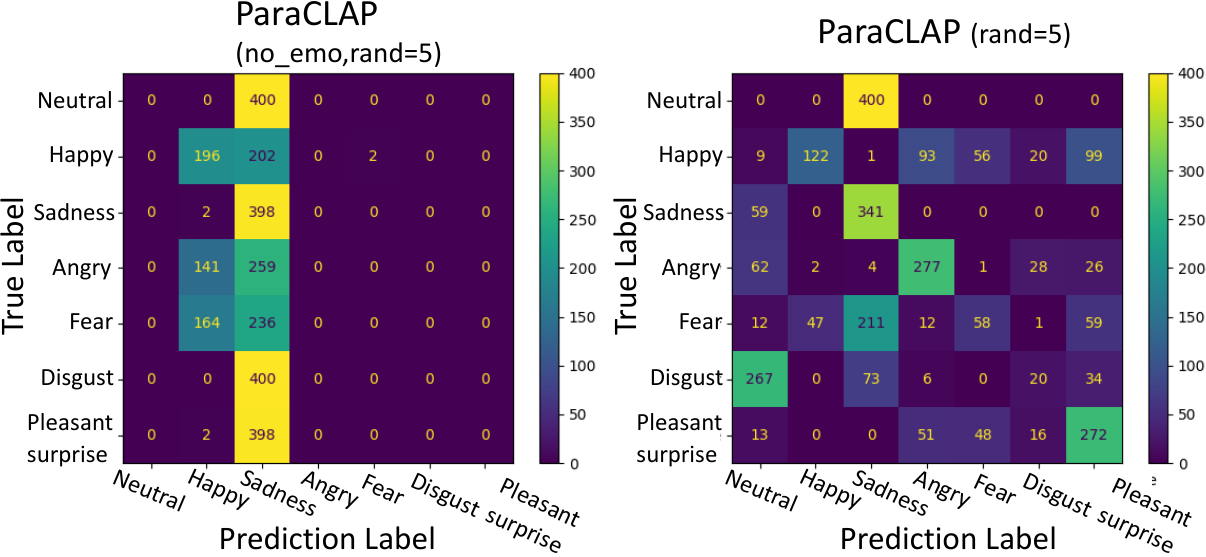}
    \caption{The confusion matrices illustrate the results on the TESS dataset. Models share identical training settings, differing in the inclusion of emotion queries within the audio-text pairs (left: without, right: with)}
    \vspace*{-4mm}
    \label{fig:cm-tess}
\end{figure}

The results are shown in \cref{tab:emods}. 
The datasets involved in the experiments are characterised by the number of classes (Class), the language they contain (Lang.), and specific training conditions, including the use of emotion queries (`emo') and concatenation of random queries.

Our first observation is that our models generally outperform both the original CLAP baseline and Pengi, showcasing how our model improves upon the state-of-the-art zero-shot method used in most recent works.
In terms of ParaCLAP alternatives, it is evident that a model only trained with emotion queries (i.\,e., ``emotion is [EMOTION]'') is showing the best overall performance.
This is especially true for datasets which show a big overlap in their emotion classes with MSP-Podcast, i.\,e., standard categorical \ac{SER} datasets.
For example, the sets of labels in MSP-Podcast and IEMOCAP are identical.
In this case, our method is tantamount to doing standard classification, and we expect the optimal performance when optimising specifically for this task.

It is only when turning to alternative \ac{SER} formulations, such as the one in FAU-AIBO, that the addition of template queries shows its strength.
This helps improve performance to a UAR of $.604$ and $.232$, over $.535$ and $.216$ for the $2$- and $5$-class tasks, respectively, compared to \eclap(only emo).
On a further positive note, this shows how our model can generalise beyond English data, even though both the upstream training of the speech encoder and CLAP fine-tuning only contained English data.
However, these gains do not translate further to tasks beyond emotions (ALC/SLD), where \eclap(only emo) still shows the best performance.

Further insight into the impact of the use of emotion labels to generate queries can be found in \cref{fig:cm-iemo} and \cref{fig:cm-tess}, which show the confusion matrix of \eclap(no emo, $rand=5$) and \eclap($rand=5$) on IEMOCAP and TESS.
In the absence of emotion queries during training, the model tends to categorise all utterances as ``sadness''\footnote{Similar behaviour is observed for the other datasets, whose confusion matrices are included as supplementary material.}.
This is partially rectified through the inclusion of emotion queries, but still remains a challenge -- especially for TESS.
We hypothesise that this is caused by a discrepancy in the type of data included in MSP-Podcast compared to the other emotional datasets considered here; MSP-Podcast is a naturalistic corpus collected `in-the-wild' whereas all four emotional datasets are acted -- and therefore contain more `archetypal' depictions of emotions (i.\,e., high arousal).
The depiction of sadness in MSP-Podcast may be less pronounced and more similar to the (overrepresented) neutral class, thus causing misclassifications towards that when the model is applied on other data.

Overall, our results show that increasing the diversity of the training queries is important to improve performance.
The inclusion of the labelled emotion in these queries additionally plays a key role -- especially when evaluating on related tasks.
This illustrates how the quest to identify good strategies for creating queries that lend themselves to generalisation across different \ac{CP} tasks is still open.

\section{Conclusions}
\label{sec:conclusion}
In this study, we conducted a preliminary investigation of different factors that can influence the performance of a CLAP-style model specifically designed for computational paralinguistics tasks.
In this process, we have created the \eclap model that can substantially outperform the state-of-the-art generic CLAP model widely used for zero-shot evaluation in recent works.
Furthermore, by exploring a templating method to generate text queries derived from handcracted features and expert knowledge, we have shown how diversity in training can be beneficial for performance.

In summary, this first attempt to create \eclap shows great promise, albeit demonstrating that there is still a lot of ground to be covered.
More diversity in the training queries is required, perhaps augmented by the use of \acp{LLM}, as done for audio captioning~\citep{Mei23-WAC}, or even a more seamless integration with \acp{LLM}~\citep{Chu23-QAA, Kong24-AFA, Deshmukh24-PAA}.
Moreover, the role of the training data still remains to be investigated, with data quantity and diversity expected to be another key factor.


\section{Acknowledgements}

\ifinterspeechfinal
     This work was funded by the China Scholarship Council (CSC), Grant \#\,202006290013, and by the DFG,  Reinhart Koselleck-Project AUDI0NOMOUS (Grant No.\ 442218748).
\else
     The authors also acknowledge the participants in the discussions related to this research, whose contributions were instrumental in refining the ideas presented. Additionally, we appreciate the support that facilitated the development of this research
\fi

\section{\refname}
\printbibliography[heading=none]

\end{document}


\section{Supplementary material -- Queries}
This appendix documents all the queries used in our work that are derived from dimensional emotion attributes and expert features.
This information is shown in \cref{tab:queries}.

\begin{table}[h]
    \centering
    \caption{
    Comprehensive list of queries generated using our templating mehtod.
    }
    \label{tab:queries}
    \resizebox{\textwidth}{!}{
    \begin{threeparttable}
    \begin{tabular}{c|ccc}
        \toprule
        \textbf{Attribute} & Low [0\%-30\%] & Mid [30\%-70\%] & Top [70\%-100\%]\\
        \midrule
         \multirow{3}{*}{Arousal} & has low arousal & arousal is at an average level & has high arousal\\
         & speaker is calm & & speaker is aroused\\
         & speaker is not very EMOTION$^\dag$ & & speaker is very EMOTION$^\dag$ \\
         
         \midrule
         \multirow{3}{*}{Valence} & has low valence & valence is at an average level & has high valence\\
         & speaker appears to be in a bad mood & & speaker appears to be in a good mood\\
         
         \midrule
         \multirow{2}{*}{Dominance} & has low dominance & dominance is at an average level & has high dominance\\
         & & & speaker appears to be dominant\\

         \midrule
         \multirow{2}{*}{Pitch ($\mu$)} & has a low pitch & has an average pitch & has a high pitch\\
         & & has a normal pitch & \\

         \midrule
         \multirow{4}{*}{Pitch ($\sigma$)} & has a low pitch variation & has a normal pitch variation & has a high pitch variation\\
         & & has a low pitch variance & has a high pitch variance\\
         & & has a very unstable pitch & has a very stable pitch\\
         & & has a very unstable phonation & has a very stable phonation\\

         \midrule
         \multirow{3}{*}{Intensity} & has a low equivalent sound level & has a normal equivalent sound level & has a high equivalent sound level\\
         & is quiet & has an average equivalent sound level & sound pressure is elevated\\
         & is almost silent & loudness is just about right & sound level is elevated\\
         & & & is loud\\

         \midrule
         \multirow{5}{*}{Duration} & has a short duration & is of average duration & has a long duration\\
         & has a small duration & is of average length & has a big duration\\
         & is a short sentence & duration is medium & is a long sentence\\
         & lasts a little time & is neither long nor short &  lasts a long time\\
         & is short & & is long\\

         \midrule
         \multirow{5}{*}{Jitter} & has a low jitter & has a normal jitter & has a high jitter\\
         & [...] but a high pitch variance$^*$ & & [...] but a low pitch variance$^*$\\
         & [...] but not a low pitch variance$^*$ & &[...] but not a high pitch variance$^*$\\
         & [...] but the pitch is unstable$^*$ & & [...] but the pitch is stable$^*$\\
         
         \midrule
         \multirow{5}{*}{Shimmer} & has a low shimmer & has a normal shimmer & has a high shimmer\\
         & [...] but a high pitch variance$^*$ & & [...] but a low pitch variance$^*$\\
         & [...] but not a low pitch variance$^*$ & &[...] but not a high pitch variance$^*$\\
         & [...] but the pitch is unstable$^*$ & & [...] but the pitch is stable$^*$\\
         \bottomrule
    \end{tabular}
    \begin{tablenotes}
    \item $^*$ Query appended depending on value of Pitch ($\sigma$)
    \item $^\dag$ Query depends on the value of the EMOTION label for that sample; only used if annotators agree to sufficient extent
    \end{tablenotes}
    \end{threeparttable}
    }
\end{table}

\section{Supplementary material -- Confusion Matrices}

\begin{figure}[H]
\centering
\subfigure[The confusion matrices illustrate the results on the RAVDESS dataset]{
\label{Fig.sub.1}
\includegraphics[width=0.45\textwidth]{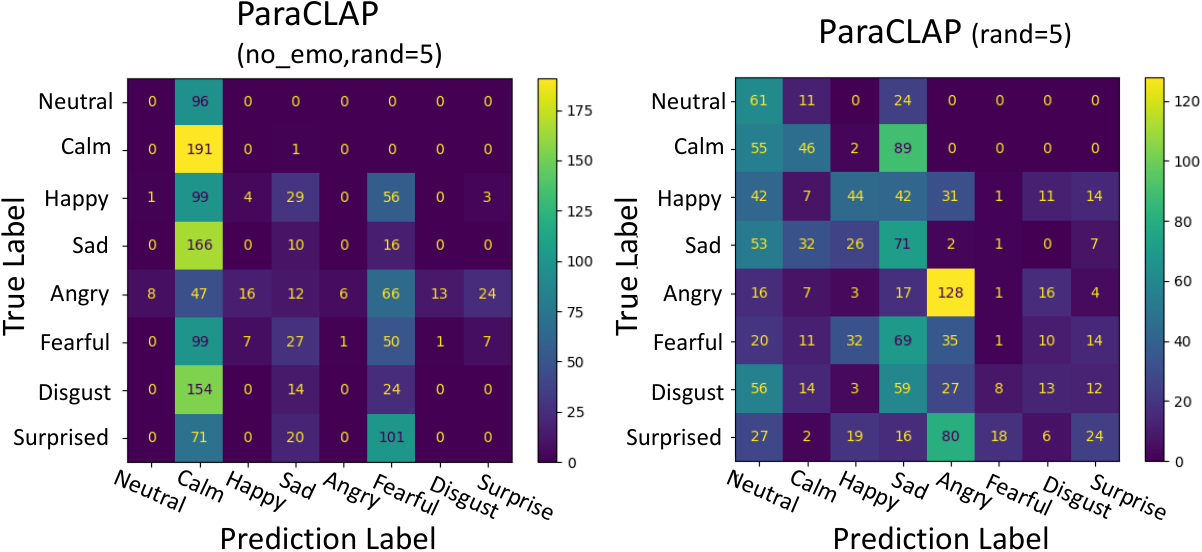}}
\subfigure[The confusion matrices illustrate the results on the CREMA-D dataset]{
\label{Fig.sub.2}
\includegraphics[width=0.45\textwidth]{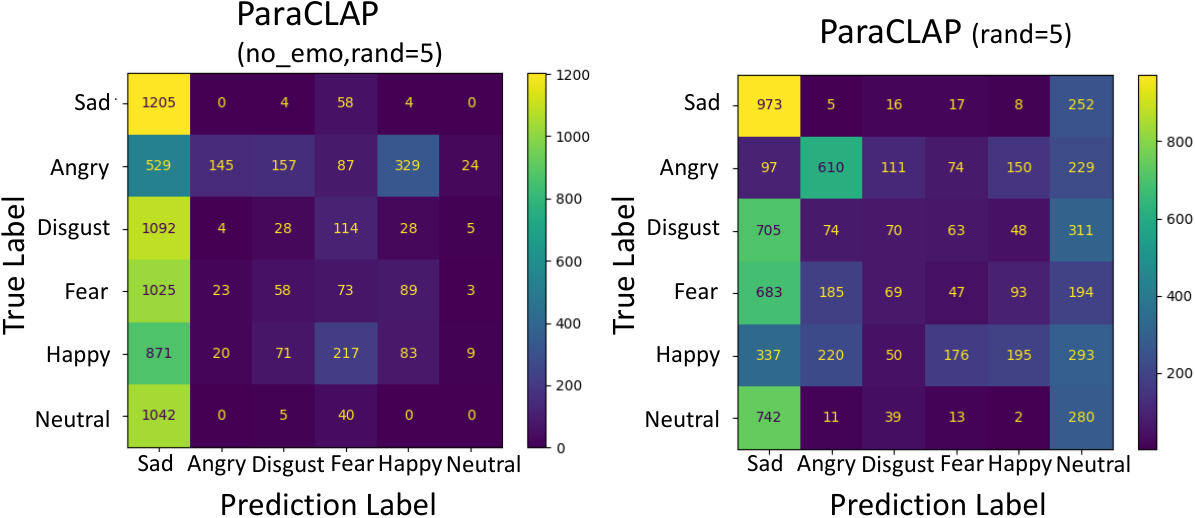}}
\subfigure[The confusion matrices illustrate the results on the FAU Aibo with 2 classes dataset]{
\label{Fig.sub.1}
\includegraphics[width=0.45\textwidth]{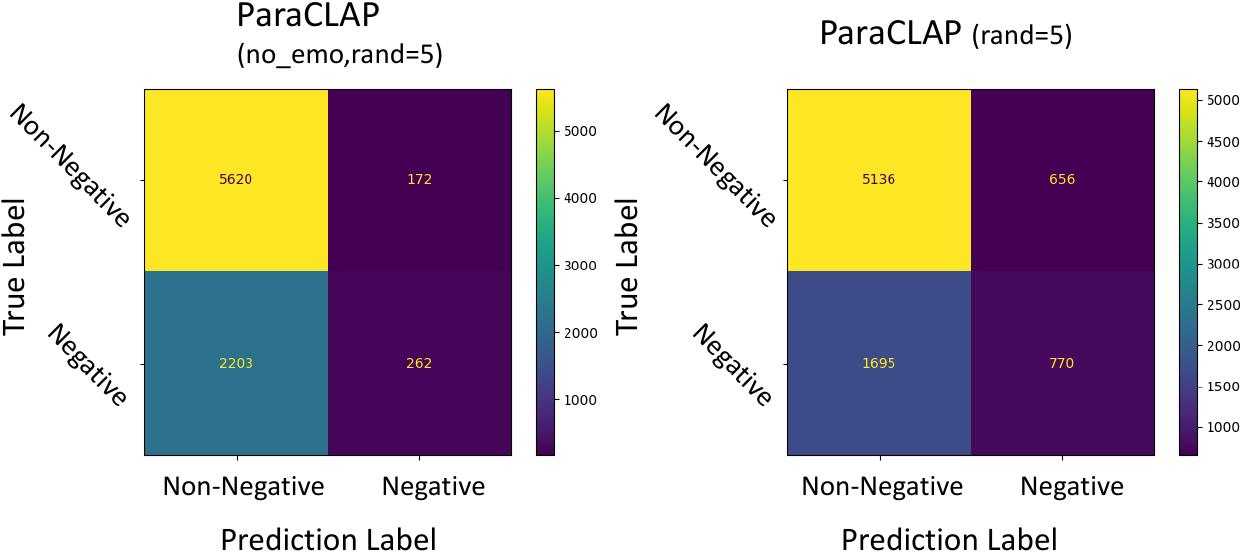}}
\subfigure[The confusion matrices illustrate the results on the The confusion matrices illustrate the results on the FAU Aibo with 5 classes dataset]{
\label{Fig.sub.1}
\includegraphics[width=0.45\textwidth]{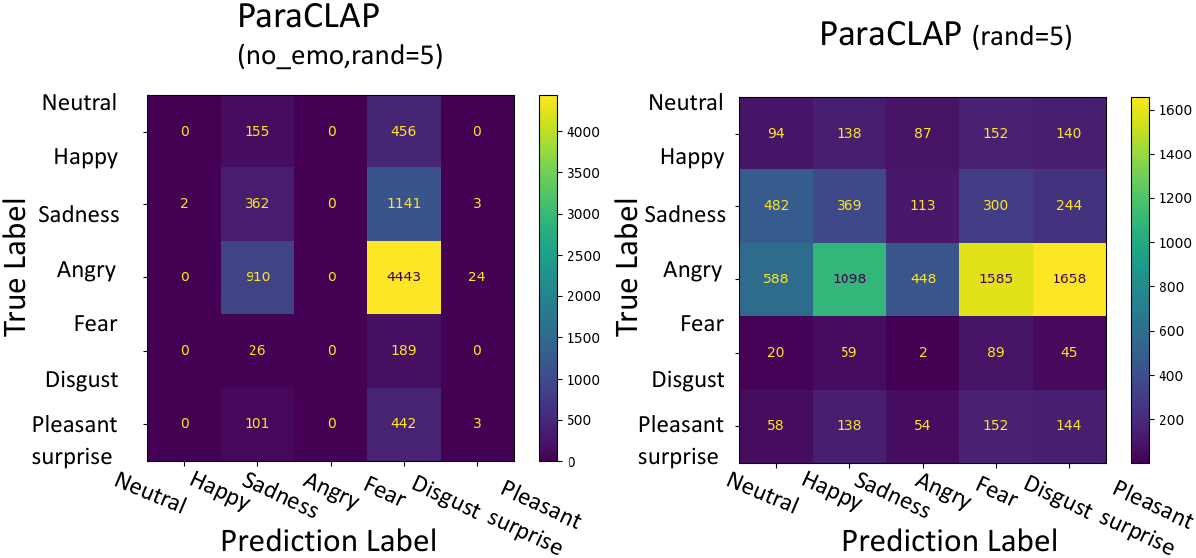}}
\subfigure[The confusion matrices illustrate the results on the ALC dataset]{
\label{Fig.sub.1}
\includegraphics[width=0.45\textwidth]{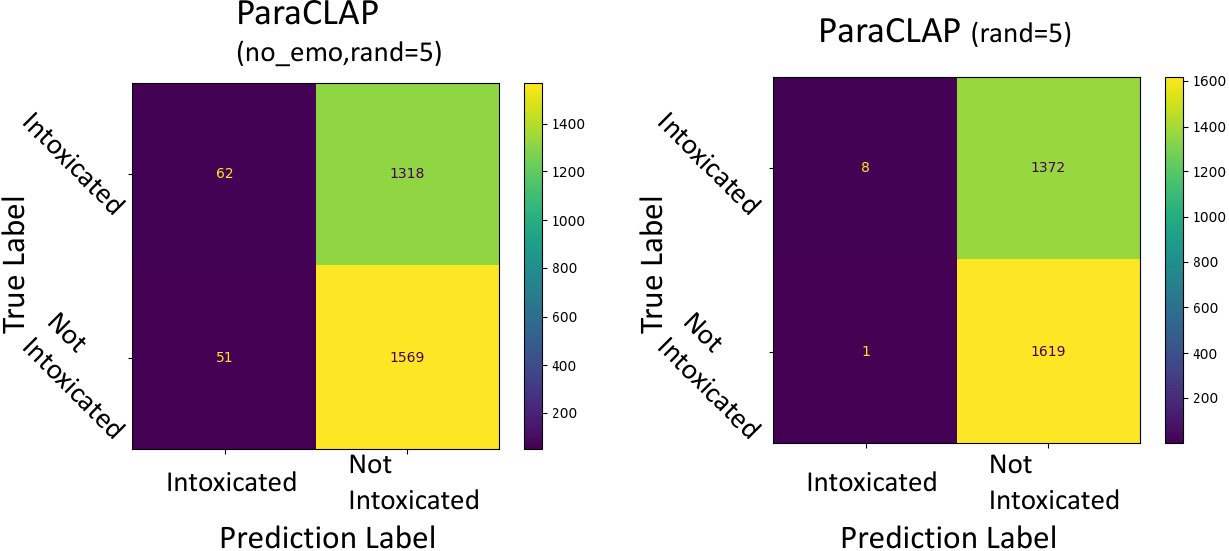}}
\subfigure[The confusion matrices illustrate the results on the SLD dataset]{
\label{Fig.sub.1}
\includegraphics[width=0.45\textwidth]{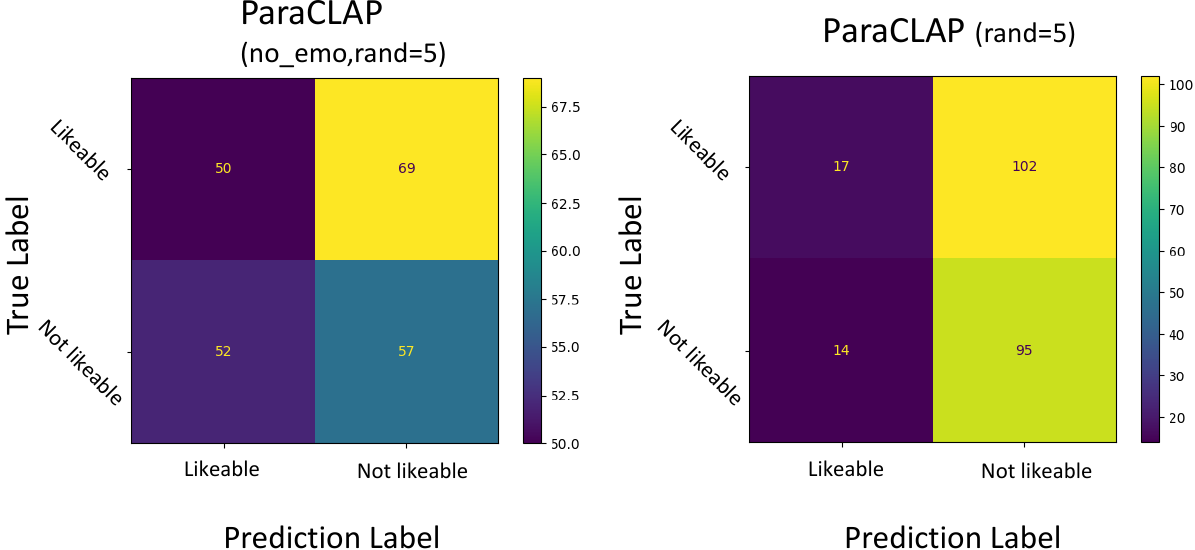}}
\caption{The confusion matrices of all other datasets. Models share identical training settings, differing in the inclusion of emotion queries within the audio-text pairs (left: without, right: with)}
\label{Fig.main}
\end{figure}